# Proposed Ultra Wide-band Tunable THz RT lasers based on momentum space negative conductivity under electron streaming in hNB-graphene sandwiches

A. A. Andronov and V. I. Pozdniakova

Institute for Physics of Microstructures RAS, Nizhny Novgorod, Russia

E-mail: andron@ipmras.ru

**Abstract.** Interpretation of recent observations by T. Otsuji group in Sendai of switching from loss to gain in THz normal incident transmission at $T = 300$ K through multiple gate systems on hNB–graphene sandwiches with rise in electric field in graphene is proposed. The observations are explained as due to THz dispersion and negative conductivity at **momentum** space transit time resonance under non equilibrium anisotropic "streaming" electron distribution in graphene under strong electron-optical phonon scattering. These phenomena are well known and established in standard semiconductors terminated by demonstration of wide band tunable microwave lasing from 50 GHz to 200 GHz by changing electric field in InP crystal at $T \approx 4$ K from 200 to 400 V/cm by L.Vorobyev group in S.Ptersburg in 2001. By generalizing these semiconductor considerations, estimate of scattering rates and with model calculation of the resonance phenomena in graphene we claim that these phenomena are responsible for the transmission results observed by T. Otsuji group with amplification at resonances determined by **spatial** transit time resonance across gates. Finally we propose the THz source without gates which consists just of intrinsic silicon piece (a resonator) placed on the sandwich with current flow in graphene providing tunable by applied electric field lasing presumably in 0.5 to 2-3-5 THz band. The source is direct THz analog of the microwave laser of L.Vorobyev group. Steps to move further this proposal are mentioned.

1. INTRODUCTION

Simple cheap tunable CW THz sources (and also detectors and mixers) working at RT (room temperature) are needed for a lot of applications starting from gaseous spectroscopy, heterodyning, observation of hidden objects etc. And recent observations of THz electromagnetic wave switching from loss to gain in the normal transmission at 300 K through multiple gated graphene–hBN sandwich under rise in electric field and current flow along graphene layer [1] open way for such sources. In that work time domain THz approach was used to study gated plasmon resonances and values of THz normal transmission through multiple gate systems with different gate widths placed on the hNB-graphene sandwiches with biased graphene. Scheme of the observations results are shown in Fig. 1 together with transit time frequency $f_E$ and optical phonon scattering parameter $M$. The gated 2-d electron gas represent an analog of the Fabry–Perot cavity produced by contra propagating gated plasmons which have high reflection on gate edges. The plasmons have constant phase velocity which depends on electron density in 2-d gas. The density in turn may be changed by DC gate voltage so that the gated plasmons represent tunable by gate voltage (and also by current flow due to the Doppler effect) Fabry–Perot cavity. Recently such tuning was demonstrated also in work [2] on tunable THz detection on gated plasmons in similar hNB- grapheme sandwiches. These two work [1,2] should be considered as originated from two old enough proposals [3,4] by Dyakonov and Shur on possibility of amplification and detection in gated plasmon system in pre graphene era. Though the THz observations in [1,2] are unique they are quite convincing. They become possible due to fabrication of graphene–hBN sandwiches which permits to have the sandwiches with graphene mobility above 50 000 cm$^2$/(V·sec) and drift velocities up to about 7·10$^7$ cm/sec at $T$ = 300 K (see e.g. [5,6]).

And the works [1,2] open way to RT THz tunable graphene electronics. But for the way to be really open it is necessary to explain mechanism of THz amplification observed in work [1] (that had not been done there). And we argue [7] that anisotropic "streaming" electron distribution in grapheme–hNB sandwiches in high electric field takes place in the experiments while the transit time resonance THz dispersion and negative conductivity effects under the streaming nearby the resonance frequency $f_E = 1/\tau_E$ (determined by time $\tau_E$ of free electron acceleration in electric field from energy $\varepsilon \approx 0$ up to energy of optical phonon $\hbar\omega_0$) explain the observation results. And basing on these considerations we propose the "ultimate" ultra wide band tunable THz RT sources using the THz negative conductivity under streaming in the sandwiches.

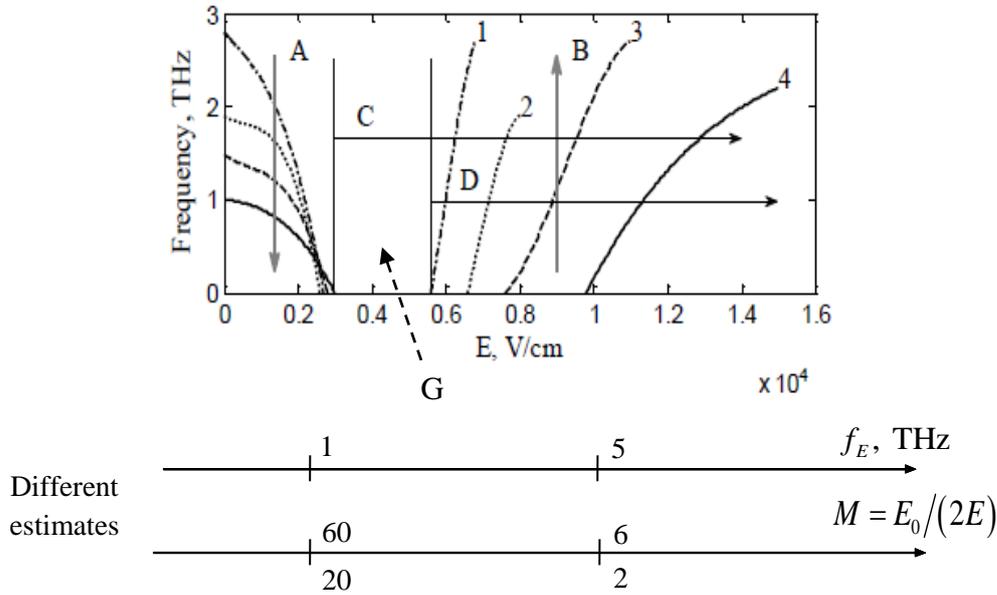

Fig. 1. Scheme of observed resonance frequencies of gated plasmon versus electric field E in graphene-hNB sandwiches adopted from [1] with gate widths ( lines): 1 – 0.5, 2 – 0.75, 3 – 1.0, 4 – 1.5 microns. Regions: A – absorption; B – amplification; C – electron streaming with negative real part of conductivity; D – emerged negative shoulder of imaginary part of conductivity; G – no resonances. Below: the transit time frequencies $f_E$ and estimates of optical phonons scattering number $M$.

2. ELECTRON STREAMING IN GPAPHENE

Streaming is cigar-like electron distribution (Fig. 2) under strong electron–optical phonon scattering at electron energy $\varepsilon$ higher than optical phonon energy $\hbar\omega_0$ and weak scattering at energy lower $\hbar\omega_0$. Under these conditions an electron accelerated at intermediate electric field reaches optical phonon energy $\hbar\omega_0$ with electron momentum $p = p_0$ almost without scattering and then quickly return back to $p \approx 0$ due to strong optical phonon emission. Weakness of scattering at energy $\varepsilon < \hbar\omega_0$ is due to low lattice temperature $T$ (kT<< $\hbar\omega_0$) so that electron scattering due to optical phonon absorption is negligible and weak scattering by acoustical phonons and impurities. If characteristic scattering rate by optical phonons at $\varepsilon > \hbar\omega_0$ is $\upsilon_0$ while characteristic scattering rate at $\varepsilon < \hbar\omega_0$ is $\upsilon_p$, and $\upsilon_0 >> \upsilon_p$, then streaming electron distribution exists at electric field $E$ within region $\upsilon_0 > eE/p_0 > \upsilon_p$. In this case an electron performed cyclic movement in momentum space with characteristic **transit time frequency** $\omega_E = 2\pi eE/p_0 = 2\pi f_E$, determined by transit (acceleration) time $\tau_E = (eE/p_0)^{-1}$ from $p \approx 0$ to $p \approx p_0$. It is the dispersion and negative conductivity around this transit time frequency represents here basis of explanation of data in [1] and the proposed THz source. The negative conductivity and dispersion effects around the transit time frequency are due to momentum space electron bunching due to joint effect of AC and DC field and optical phonon emission. The bunching is alike that in semiconductor superattices [8, 9] and spatial electron bunching in vacuum electronic sources. Streaming is well known in semiconductors [10, 11] as well as the high frequency effects and negative conductivity around the transit time frequency under streaming. The latter were proposed [12,13], simulated at microwaves in GaAs [14] and InP [15], at THz in GaN [16] and demonstrated in [17] with tunable microwave lasing from 50 to 200 GHz in InP crystal by changing applied electric field from 200 to 400 V/cm.

In graphene and hNB optical phonon energy is about 0.2 eV and electron coupling with the optical phonons is strong enough. So in high quality graphene–hNB sandwiches condition needed for streaming can be well satisfied even at 300 K. Recent Monte-Carlo simulations of electron distribution in single graphene layer in high electric field performed in [18] demonstrate weak streaming. Form the cigar in graphene due to Dirac momentum $p$ energy $\varepsilon$ dependence ($\varepsilon = V_F p$, $V_F$ is constant so call Fermi velocity) is different from the one in semiconductors (see e.g. [11]). In graphene 2-d electron density of state is proportional to p. In this case (if we take as in [18] constant deformation potential for electron – optical phonon interaction) electron scattering rate due to spontaneous optical phonon emission $\upsilon_{opt}$ can be written as $\upsilon_{opt} = \upsilon_0 (p - p_0)/p_0$. Now if we suppose (as in [11]) that there is no scattering at $\varepsilon < \hbar\omega_0$ and only spontaneous emission of optical phonons by electrons exist the Boltzmann equation for momentum space distribution function $F(\boldsymbol{p}) = F(p_x, p_z)$ with electric field $\boldsymbol{E}$ along z-axis has the following form.

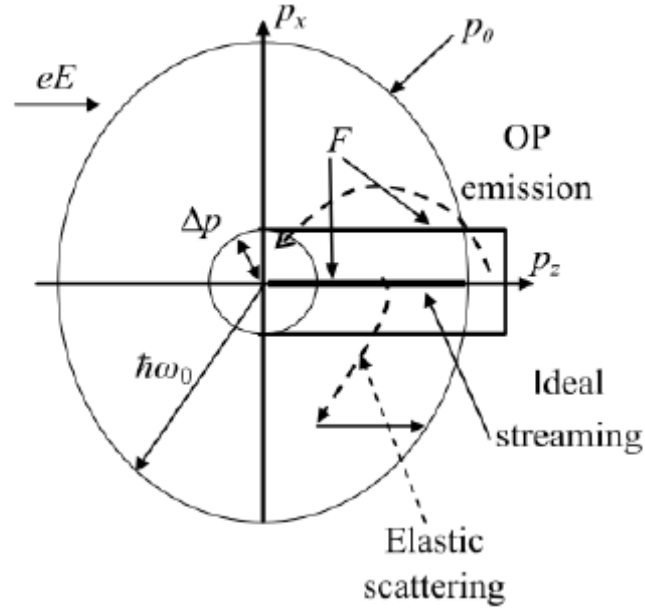

Fig. 2. Schemes of electron distributions under streaming. Lines of constant electron distribution $F$ for realistic and ideal streaming; OP – optical phonon emission

In the active region (for $\varepsilon > \hbar\omega_0$) with $F(\boldsymbol{p}) = F_a(\boldsymbol{p})$:

$$eE \frac{df_a}{dp_z} = -F_a(p) \frac{\upsilon_0 (p - p_0)}{p_0}. \tag{1}$$

For small value of electric field $E \ll E_0 = \upsilon_0 p_0/e$ – characteristic optical phonon scattering field approximate expression for $F_a(p)$ is:

$$F_a(p) = F(p_{z0}) \exp\left(-M \frac{(p-p_0)^2}{p_0^2}\right), \qquad (2)$$

$F(p_{z0}) = F(p_x, p_{z0})$ is distribution function at $p = p_0$, $M = \upsilon_0 p_0/(2eE) = E_0/(2E)$. This parameter $M$ is presented in Fig. 1 with two performed below estimates for $E_0$. From (2) we get the characteristic momentum penetration depth to $\varepsilon > \hbar\omega_0$

$$\langle \Delta p_z \rangle_{av} \approx p_0 \sqrt{\frac{2E}{E_0}} = \frac{p_0}{\sqrt{M}}. \qquad (3)$$

In the passive region (for $\varepsilon < \hbar\omega_0$) $F(\mathbf{p}) = F_p(\mathbf{p})$ and the Boltzmann equation is

$$eE \frac{dF_p}{dp_z} \approx \frac{\upsilon_0}{2\pi p_0} \exp\left(-M \frac{p^2}{p_0^2}\right) \int F_a(p_x, p_{z0}) dp_x = I \exp\left(-M \frac{p^2}{p_0^2}\right). \qquad (4)$$

Now

$$F_p = \frac{1}{eE} I \int_{-\infty}^{p_z} \exp\left(-M \frac{p_x^2 + p_z^2}{p_0^2}\right) dp_z$$

and for $p_x \gg p_0/\sqrt{M}$

$$F_p = \frac{I}{eE} \sqrt{\frac{\pi}{M}} \exp\left(-M \frac{p_x^2}{p_0^2}\right) \qquad (5)$$

Now we see that both the width of the streaming cigar and its penetration depth in electric field beyond $p_0 = \hbar\omega_0/V_F$ are the same and equal to $\Delta p = p_0/\sqrt{M}$ as shown in Fig. 2. We will use the formula (5) in calculation of model THz conductivity under streaming in graphene.

We show below that in the observation data shown in Fig. 1 streaming exist in the drift velocity region indicated and that momentum space transit time dispersion and negative conductivity phenomena are responsible for the observation. The momentum space transit time frequencies $f_E = \omega_E/2\pi \approx 1.65 - 5$ THz shown in Fig. 1 in the region of electric fields (3–10 kV/cm) is just the region close to the observed plasmon resonance frequencies where shift to THz amplification in transmission was observed. This is one of the argument for the interpretation of the observed in [1] phenomena in the framework of streaming and transit time resonance.

Another argument to support possibility for occurrence of streaming and the transit time phenomena in hNB–graphene sandwiches at 300 K comes from estimate of scattering rates in the sandwiched. To estimate the rates we take simplest approach used in work [18] with constant deformation potential for optical and acoustical phonons scattering and charged impurity scattering without screening by electrons according to [19–21]. Scattering rates for optical $\upsilon_{opt}$ and acoustical $\upsilon_{ac}$ phonons scattering and for impurity scattering $\upsilon_{imp}$ may be written as:

$$\upsilon_{opt} = \upsilon_0 \frac{p - p_0}{p_0}, \quad \upsilon_{ac} = \upsilon_{a0} \frac{p}{p_0}, \quad \upsilon_{imp} = \upsilon_{i0} \frac{p_0}{p}. \qquad (6)$$

For optical phonon scattering (as said above) only scattering due to spontaneous optical phonon emission is included. Scattering rate $\upsilon_p$, which determines condition providing streaming should be considered as averaged over streaming distribution and can be written as

$$\upsilon_p = \frac{\upsilon_{a0}}{2} + \upsilon_{i0} \ln \frac{p_0}{p_{min}} \qquad (7)$$

where $p_{min}$ – cutting parameter, which can be taken as width of streaming $\langle \Delta p_x \rangle_{av} = p_0/\sqrt{M}$. Now

$$\upsilon_p = \frac{\upsilon_{a0} + \upsilon_{i0} \ln M}{2} = \upsilon_{pa} + \upsilon_{pi}. \qquad (8)$$

With parameters used in [18] for single graphene layer we have $\upsilon_0 = 10^{13}$ sec$^{-1}$, $\upsilon_{a0} = 10^{12}$ sec$^{-1}$ (for 300 K). So $\upsilon_0/\upsilon_{pa} \approx 20$, while $E_0 \approx 20$ kV/cm and $E_p \approx 1$ kV/cm. This figures come from consideration of single grapheme layer. But in the sandwiches graphene is encapsulated in two hNB layers which have similar to grapheme lattice parameters. And surface phonons in the hNB layers should contribute to phonons scattering rates. We are unaware of the scattering rates for the sandwiches. But keeping in mind the similarity we can triple the phonons scattering rates to get $\upsilon_0 = 3 \cdot 10^{13}$ sec$^{-1}$, $\upsilon_{a0} = 3 \cdot 10^{12}$ sec$^{-1}$ and $E_0 \approx 60$ kV/cm and $E_p \approx 3$ kV/cm. The last value (3 kV/cm) is just the value where plasmon resonance disappeared (Fig. 1) due to beginning of streaming as argued below.

Charge impurity scattering in graphene calculated e.g. in [19–21]. Without screening one has

$$\upsilon_{imp0} = N_{imp} \frac{\pi^2 e^4}{\hbar^2 \kappa^2 \omega_0}. \qquad (9)$$

Here $N_{imp}$ (1/cm$^2$) – impurity concentration, $\kappa$ – dielectric permittivity (in hNB). If we take $N_{imp} = 10^{10}$ cm$^{-2}$, $\kappa = 6.5$ [1] we get $\upsilon_{imp0} = 3.3 \cdot 10^{11}$ sec$^{-1}$, i.e. of about that $\upsilon_{pa}$. This means that the electric field region where streaming can exist at 300 K is in the range 1–20 kV/cm or more likely 3–60 kV/cm. This is the range show in Fig. 1. It should be stressed that impurity concentration $N_{imp}$ has nothing to do with electron concentration $N$ in graphene in work [1]. The concentration $N$ was fixed by value of the gated voltage while $N_{imp}$ originates from sandwich fabrication.

The above discussions we think clearly demonstrate that in high quality hNB–graphene sandwiches it is possible to achieve at 300 K electron streaming and that streaming takes place in work [1]. Now we will discuss THz transit time phenomena under streaming in graphene.

### 3. THz Dispersion and NDC under Streaming and Momentum Space Transit Time Effects in Gpaphene

As mentioned above under streaming there exists transit time frequency $\omega_E = 2\pi eE/p_0 = 2\pi/\tau_E = 2\pi f_E$, where $\tau_E$ is acceleration time in electric field $E$ from $p = 0$ to $p = p_0$. Consider now streaming response to small AC electric field $\tilde{E} = \tilde{E}_0 \exp(i\omega t)$ parallel to DC field $E$ at frequencies $\omega \approx \omega_E$, i.e. near the transit time resonance. The response is determined by electron bunching in momentum space and depends on processes which define width of the resonance. The latter are due to spread of electron transit times as a result of final time of optical phonon emission $\tau_0 = 1/\upsilon_0$ [13] or elastic electron scattering at $p < p_0$ [15,17].

Let us first discuss the electron response nearby $\omega_E$ for semiconductors with constant electron mass to establish background for discussion in graphene. In work [13] the response was calculated in a semiconductor with constant electron mass for finite but small optical phonons scattering time $\tau_0 = 1/\upsilon_0^*$ ($\upsilon_0^*$ is similar to introduced above characteristic optical phonon scattering rate) without elastic scattering at $\varepsilon < \hbar\omega_0$. Small $\tau_0$ provided small penetration of electrons in DC field beyond $\hbar\omega_0$ and small spread of electron transit times. For high dimensionless transit time frequency $\Omega = \omega \tau_E$ and small $\tau_0$ calculated conductivity of electron system is:

$$\sigma = \sigma_\Omega = \frac{\omega_p^2 Q(x)}{4\pi \omega_E \Omega} \cdot \frac{-\sin\Omega + i(R - \cos\Omega)}{(R - \cos\Omega)^2 + R^2 \sin^2 \Omega}. \qquad (10)$$

Here $\omega_p$ is plasma frequency of electrons, $Q(x) = 0.1x^2$, $R = 1 - 0.15x^2$, $x = \omega(\tau_0 \tau_E)^{1/3} \ll 1$. Plot of dimension less conductivity $\Sigma_\tau^P(\Omega) = \sigma_\Omega \left(\omega_p^2/(4\pi\omega_E \Omega)\right)^{-1}$ for $x = 0.5$ is presented in Fig. 3.

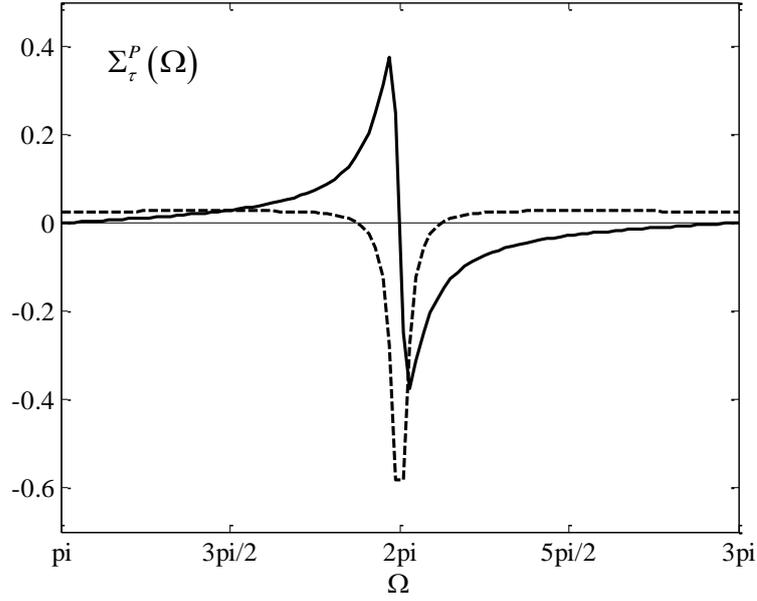

Fig. 3. Dimension less conductivity $\Sigma_\tau^P(\Omega)$ for $x = 0.5$. Solid line – $\text{Re}(\Sigma_\tau^P(\Omega))$, broken line – $\text{Im}(\Sigma_\tau^P(\Omega))$.

We see that the response demonstrates oscillating form of real part of conductivity with change to negative conductivity at $\Omega > 2\pi$ ($\omega > \omega_E$) and negative dip of imaginary part at $\Omega = 2\pi$ ($\omega = \omega_E$). The negative dip of the imaginary part for used AC electric field $\tilde{E} = \tilde{E}_0 \exp(i\omega t)$ imply that the system for appropriate plasma frequency can support plasma oscillations at $\omega \approx \omega_E$ which can be excited by real part negative conductivity at $\omega > \omega_E$. Such situation was simulated in [22].

Such behavior of the conductivity is typical for active dissipative systems where real and imaginary parts change places compared with standard active or dissipative two level systems. Similar situation is for conductivity in semiconductor superlattices in electric field around the Bloch frequency. Here real part of conductivity also change sign at resonance (at the Bloch frequency) while imaginary part has peak at resonance [8, 23]. But this peak (for AC electric field $\tilde{E} = \tilde{E}_0 \exp(i\omega t)$ – this field was used in [8]) is **positive** because real part is negative at frequencies **below** the resonance. Surely it is in accord with Kramers-Kronig relations. And the plasmons excitation is impossible. We will see below that at the beginning of streaming in graphene the conductivity behavior is alike that in superlattice and plasmons do not exist. On the other hand at higher electric field situation is more complicated and interesting.

Though the results presented in (10) and the Monte-Carlo simulation with similar results for real part of conductivity under streaming performed in works [14–16] (as well as experiments in [17]) refer to a semiconductor with constant effective mass demonstrate common nature of the dispersion and negative conductivity effects under streaming. The final conductivity appeared here with non zero optical phonon scattering time $\tau_0$ or due to (small) elastic scattering rate $\upsilon_p$ at $\varepsilon < \hbar\omega_0$ (cf. [15,17]).

To demonstrate effect of (small) elastic scattering rate $\upsilon_p$ at $\varepsilon < \hbar\omega_0$ at the response it is possible to consider the limiting case with $\tau_0 = 0$ which permits simple consideration of transit time frequencies and resonances, and perform model calculation of conductivity nearby $\omega_E$. First we perform the latter for semiconductor with constant effective mass. Then we move such a treatment to graphene which permits to see general picture of transit time resonance phenomena there and present interpretation of observations results in work [1].

For $\tau_0 = 0$ we have "ideal" streaming (Fig. 2) and for weak elastic scattering ignoring small electron occupation outside ideal streaming we can use 1-d electron distribution function $f_0(p_z)$ to calculate transit time conductivity under streaming with only "out" term due to elastic scattering in the Boltzmann equation. Now the equation for 1-d time independent $f_0(p_z)$ in constant electric field $E$ is

$$eE\frac{df_0}{dp_z} = -\upsilon f_0(p_z) + I\delta(p_z), \qquad (11)$$

where $\delta(p_z)$ is delta function while $I\delta(p_z)$ is "source" due to transitions after optical phonon emission. For used 1-d approximation $I = eEf_0(p_0)$.

The equation for amplitude $\tilde{f}_0(p_z)$ of small perturbation of the distribution function $\tilde{f}(p_z) = \tilde{f}_0(p_z)\exp(i\omega t)$ in small AC field $\tilde{E} = \tilde{E}_0 \exp(i\omega t)$ is

$$i\omega \tilde{f}_0 + eE\frac{d\tilde{f}_0}{dp_z} = -\upsilon\tilde{f}_0 + e\delta(p_z)\left(E\tilde{f}_0(p_0) + \tilde{E}_0 f_0(p_0) - \tilde{E}_0 f_0(0)\right) - e\tilde{E}_0\frac{df_0}{dp_z}. \qquad (12)$$

Here according to (11) $f_0(p_z) = f_0(0)\exp(-\upsilon p_z/(eE_0))$, $f_0(p_0) = f_0(0)\exp(-\upsilon p_0/(eE_0))$ and for $\upsilon p_0)/eE \ll 1$ $f_0(0) \approx N_0/p_0$ and (as in (11)) terms with $\delta(p_z)$ are due to (instantaneous as $\tau_0 = 0$) transition after optical phonon emission. By introducing dimension-less functions and parameters

$$x = \frac{p_z}{p_0}, \quad \varphi_0(x) = \frac{f_0(x)}{f_0(0)}, \quad \varphi(x) = \frac{\tilde{f}_0(x)}{f_0(0)}, \quad \Omega = \frac{\omega p_0}{eE_0}, \quad \tilde{\Omega} = \frac{(\omega - i\upsilon)p_0}{eE_0}$$

we get

$$i\tilde{\Omega}\varphi(x) + \frac{d\varphi}{dx} = \delta(x)\left(\varphi(1) + Q\varphi_0(1) - Q\right) - Q\frac{d\varphi_0}{dx}, \qquad (13)$$

were $\varphi_0(1) = \exp(-\mu)$, $\mu = \upsilon p_0/(eE_0)$. For supposed $\mu \ll 1$ we have

$$i\tilde{\Omega}\varphi(x) + \frac{d\varphi}{dx} = \delta(x)\left(\varphi(1) - \mu Q\right) + \mu Q \qquad (14)$$

Without AC electric field ($Q = 0$) solution of (14) gives $\varphi(x) = \varphi(1)\exp(-i\tilde{\Omega})$ and $\varphi(1) \neq 0$ if $\exp(-i\tilde{\Omega}) = 1$ or if $\tilde{\Omega} = 2\pi n$, or in another terms for

$$\omega = \omega_n = \frac{2\pi n}{\tau_E} + i\upsilon = n\omega_E + i\upsilon. \qquad (15)$$

These are the transit time resonance frequencies with decaying part due to elastic scattering.

Equation (14) with $Q \neq 0$ can be easily solved and if we take only the solution part with high $\Omega$ (as in (11)) and around resonance at $\omega_n$ we get this ("cut") expression:

$$\varphi(x) = \varphi_{cut}(x) = -\mu Q \frac{\exp(-i\tilde{\Omega}x)}{1 - \exp(-i\tilde{\Omega})}. \qquad (16)$$

This expression has resonant denominator at transit time resonance frequencies $\omega_n$; only this frequency region we are discussing. Now we can find amplitude $\tilde{J}_0$ of electron current $J_z = \tilde{J}_0 \exp(i\omega t)$ and conductivity $\sigma = \sigma_1 + i\sigma_2$:

$$\tilde{J}_0 = e \int \tilde{f}_0 v_z(p_z) dp_z = e f_0(0) \int v_z(p_z) \varphi(x) dp_z = e f_0(0) p_0 \int v_z(p_z) \varphi(x) dx$$
$$= e p_0 \frac{N_0}{p_0} \int v_z(p_z) \varphi(x) dx = \sigma \tilde{E}_0. \tag{17}$$

Here $v_z(p_z)$ is an electron velocity. We consider two cases: electrons with constant effective mass $m$ with $v_z(p_z) = p_z/m$ and graphene with $v_z(p_z) = V_F p_z/p$, $V_F$ is the Fermi velocity. For the first case calculation is simple and we have (once again with contribution from only resonance term)

$$\sigma = \sigma_\upsilon^P = \frac{e^2 N_0 p_0}{em E_0 \Omega} \Sigma_\upsilon^P(\mu, \Omega),$$

$$\Sigma_\upsilon^P(\mu, \Omega) = \mu \exp(-\mu) \frac{-\sin\Omega - i(\cos\Omega - \exp(-\mu))}{1 + \exp(-2\mu) - 2\exp(-\mu)\cos\Omega}. \tag{18}$$

The behavior of $\Sigma_\upsilon^P(\mu, \Omega)$ (Fig. 4) is well in line with $\Sigma_\tau^P(\Omega)$ (Fig. 3): conductivity real part is negative at $\Omega > 2\pi$ ($\omega > \omega_E$) while imaginary part has negative dip at $\Omega = 2\pi$ ($\omega = \omega_E$). This feature demonstrate that the conductivity behavior does not depends on electron bunching mechanism: finite $\tau_0$ or finite elastic scattering $\upsilon$. The behavior depends on electron velocity dependence on momentum as it is seen in calculated graphene conductivity.

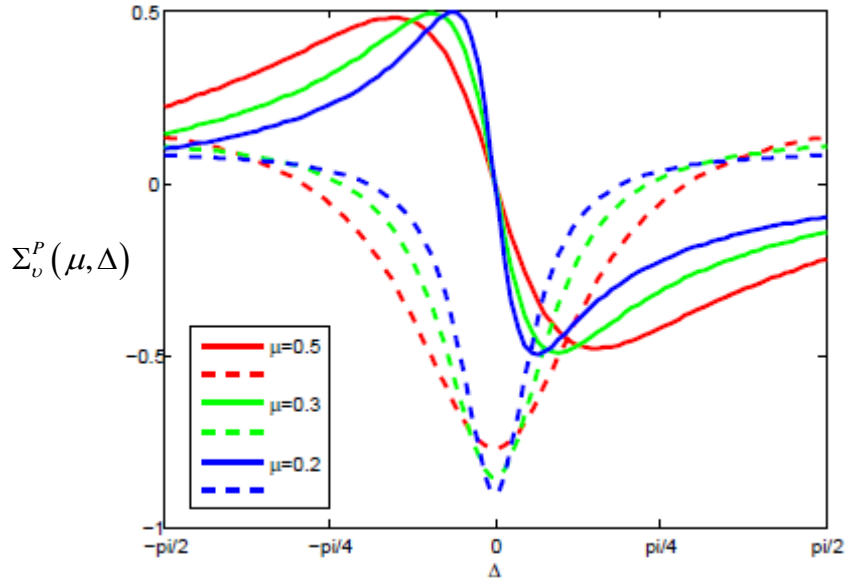

Fig 4. Dimension-less conductivity $\Sigma_\upsilon^P(\mu, \Delta) = \Sigma_1^P + i\Sigma_2^P$ for ideal streaming in parabolic band with elastic scattering at $p < p_0$ near transit time resonance $\Omega = 2\pi$, $\Delta = \Omega - 2\pi$; solid lines $\Sigma_1^P$, broken lines $\Sigma_2^P$.

In graphene $v_z(p_z) = V_F p_z/\sqrt{p_x^2 + p_z^2}$ and for ideal streaming the whole distribution function $F_0(p_x, p_z)$ and $\tilde{F}_0(p_x, p_z)$ are proportional to $\delta(p_x)$ and in $v(p_z)$ we must put $p_x = 0$ so that $v(p_z) = V_F$. With such $v(p_z)$ calculation of $\sigma = \sigma_\upsilon^D$ with $\varphi_{cut}(x)$ gives $\sigma_\upsilon^D = i\mu e^2 N_0 V_F/(eE\tilde{\Omega})$, i.e. the resonance disappeared. To "return" the resonance we should take into account width of streaming over $p_x$ by averaging expression of conductivity with $v_z(p_z) = V_F p_z/\sqrt{p_x^2 + p_z^2}$ over $p_x$ via distribution (5) or with distribution $W(p_x) = \sqrt{M/\pi} \exp(-M p_x^2)$, $\int W(p_x) dp_x = 1$. Now we get (see also [7])

$$\sigma_\upsilon^D = \frac{e^2 N_0}{\tilde{m}\omega} \Sigma_\upsilon^D(\mu, M, \Omega), \quad \tilde{m} = \frac{p_0}{V_F},$$

$$\Sigma_\upsilon^D(\mu, M, \Omega) = \frac{\mu\Omega(\exp(-\mu + i\Omega) - 1)}{1 + \exp(-2\mu) - 2\exp(-\mu)\cos\Omega} \times \qquad (19)$$

$$\sqrt{\frac{M}{\pi}} \int_0^1 x \exp\left(\frac{Mx^2}{2} - (\mu + i\Omega)x\right) K_0\left(\frac{Mx^2}{2}\right) dx.$$

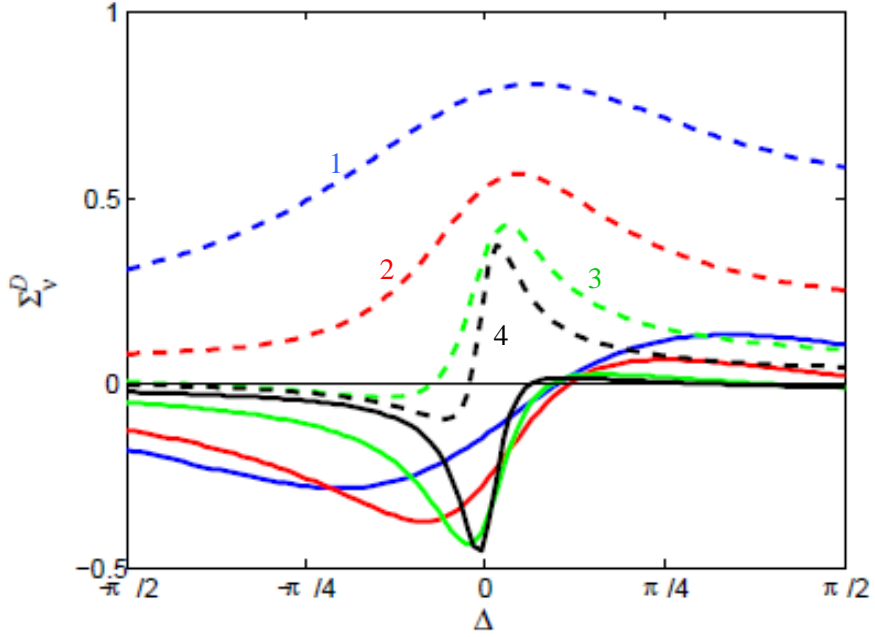

Fig. 5. Dimension-less conductivity $\Sigma_\upsilon^D(\mu, M, \Delta) = \Sigma_1^D + i\Sigma_2^D$ ($\Delta = \Omega - 2\pi$) versus dimension-less frequency under streaming according to (19); solid lines $\Sigma_1^D$, broken lines $\Sigma_2^D$; 1 – $\mu = 1$, $M = 20$; 2 – $\mu = 0.5$, $M = 10$; 3 – $\mu = 0.2$, $M = 4$; 4 – $\mu = 0.1$, $M = 2$.

Plots of $\Sigma_\upsilon^D(\mu, M, \Omega)$ are presented in Fig. 5. The plots behavior is quite different from the ones in Figs. 3, 4: real part negative conductivity is at $\Omega < 2\pi$ ($\omega < \omega_E$) at any $\mu$ and $M$ with mostly positive peak in imaginary part. And only at small $\mu$ and $M$ "shoulders" of negative imaginary part appear. Though results presented in Fig. 5 are based on very simple (perhaps even primitive) calculations they are able to explain results of work [1] shown in Fig. 1. And as such similar conductivity plots seems should emerge from more sophisticated calculations. Indeed at small field before streaming appears gated plasmons are absorbed by elastic scattering. At about 3 kV/cm streaming appears with real part negative conductivity and broad frequency region with positive imaginary part. The latter prevents existence of plasmons (as said above about the case of semiconductor superlattices) so that no resonance absorption or amplification takes place with THz transmission in large enough region of electric field (as observed in [1]).

With further rise in field $\mu$ and $M$ go down and at $\Omega < 2\pi$ the negative shoulders in conductivity imaginary part emerge. However value of the shoulder is lower than value of negative real part of conductivity. In such situation it is difficult to discuss behavior of system in terms of gated plasmons. It is more appropriate to consider this case as the one with homogeneous AC electric field under a gate alike the one in parallel plate capacitor. Still in work [1] resonances with amplification are observed in this field region just at frequencies below $f_E$ where $\mathrm{Re}\,\Sigma_\nu^D$ supposing is negative. Because the resonance amplification thresholds depends on width of the gates (Fig. 1) and electric field $E$ also determined electron drift velocity $V_D$ it is natural to suppose that the resonances are due to **spatial** electron transit through (under) gates with resonance frequency $f_S \approx V_D/d$ ($d$ is gate width). According to [1] in the streaming region drift velocity $V_D$ changes from $1.5 \cdot 10^7$ to $7 \cdot 10^7$ cm/sec and at the beginning of the resonance amplification for $d = 0.5$ microns velocity $V_D$ is $2.75 \cdot 10^7$ cm/sec. Calculated spatial transit time frequency $f_S$ with the latter value for $V_D$ is well in accord with measured value of the resonance frequency of 0.5 THz (Fig. 1). For more wide gates for the same resonance frequency 0.5 THz higher drift velocities are needed. It is just that was observed. It should be mentioned that even when simple conductor drifts through (under) the gate (parallel plate capacitor) the capacitor negative resistance can takes place as demonstrated in [24] nearby the **spatial** transit time frequency $f_S$. No doubt that similar situation could occur with the graphene conductivity under streaming. But discussion of these things required substantially more detailed consideration and calculation and is out of the scope of the present paper.

From the topic of the present paper, THz sources, the most important is existence of THz negative conductivity and amplification throughout ultra wide frequency band presumably from 0.5 THz to 3-5 THz in electric field $E > 3$ kV/cm (Fig. 1). The THz lasing due to $\Sigma_1^D < 0$ can take place even for a system with no plasmon (alike the one with microwave lasing in InP crystal as observed in [17]), say in Si resonator with whispering gallery modes placed on graphene–hNB sandwich with biased graphene (Fig. 6). The mode tails enter graphene and due to graphene negative conductivity are amplified. In this case sign of imaginary part of conductivity play no role because its contribution to mode frequency is negligible. The gate shown in Fig. 6 serves for arranging electron density in graphene suitable for the lasing.

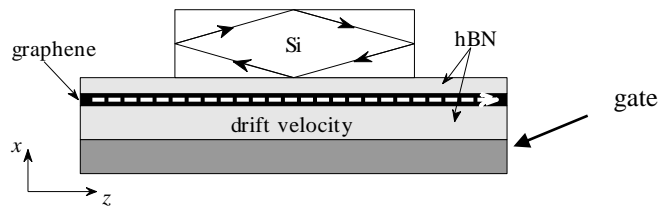

Fig. 6. Scheme of proposed THz source with Si resonator placed on graphene–hNB sandwich with biased graphene and resonator modes excitation due to transit time THz negative conductivity under streaming in graphene.

### 4. Conclusion

Interpretation of observation results of THz normal transmission at 300 K through systems of gated plasmons at hNB–graphene sandwiches with loss to gain changing at rise in electric field and current in graphene [1] is presented. This has been done via demonstration (by scattering rates estimate and model calculation) existence in the system in high electric field electron streaming with THz dispersion and negative conductivity around **momentum** space transit time resonance frequency. This approach permits to explain disappearance of plasmon resonances in a broad region of biased electric fields. At higher electric fields (outside the no resonance region) where resonances with amplification was observed in [1] also **spatial** transit effects under electron drift through (under) gate are added for the interpretation. Basing on this discussion we propose "Universal ultimate" ultra wide band (0.5–3–5 THz) tunable THz sources. To further promote this consideration more sophisticated transport and conductivity simulation are needed especially to find out limiting achievable THz frequencies both for momentum space and spatial transit time frequencies. Also development and advancing of the sandwich fabrication technology is first of all required. To study electron streaming and THz dispersion and loss/gain change observation the normal THz transmission through high quality hNB–graphene sandwich or just graphene on hNB even without gates under current flow in graphene is proposed as the next step.

ACKNOWLEDGMENTS

The authors are indebted to Vladimir Gavrilenko (IPM RAS) who put authors attention to works [1,2], to Victor Ryzhii and Vladislav Kurin for discussion and comments.
This work was done under IPM RAS Goscontract № 0035-2019-0021-C-01.